\documentclass[11pt]{article}
\usepackage{graphicx}
\begin{document}

{Inhomogeneous and self-organised temperature in Schelling-Ising model}
\bigskip

Katharina M\"uller, Christian Schulze and Dietrich Stauffer*

\bigskip
Institute for Theoretical Physics, Cologne University

D-50923 K\"oln, Euroland 

* also IZKS, Bonn University

\bigskip
Abstract: 

The Schelling model of 1971 is a complicated version of a square-lattice
Ising model at zero temperature, to explain urban segregation, based on the 
neighbour preferences of the residents, without external reasons. Various 
versions between Ising and Schelling models give about the same results. 
Inhomogeneous ``temperatures'' $T$ do not change the results much, while a 
feedback between segregation and $T$ leads to a self-organisation of an 
average $T$.

\bigskip
\section{Introduction}
Racial segregation in US cities was studied in 1971 by Schelling \cite{schelling}
through a complicated zero-temperature Ising model. This original model fails to
give large domains or ``ghettos" but later work showed realistic long-range 
order \cite{jones,ortmanns,schulze,kirman,solomon}, by introducing some 
randomness into the process e.g. via a positive temperature (= tolerance). 
The present work introduces into the standard two-dimensional Ising model an 
inhomogeneous temperature, selected randomly 
and independently for every different site, and also a feedback which increases
(decreases) the local temperature if the correlations there are too strong (too weak).

These models use two groups A and B of people (or more than two \cite{schulze}),
distributed on a square lattice, and assume that everybody prefers to be 
surrounded by lattice neighbours from the same group instead of from the 
other group. The question is whether large ``ghettos'' are formed, without
any outside intervention, where nearly everybody belongs to one group in one
residential district (domain) and to the other group in the other domain.
In the original Schelling model, unhappy people move to the closest empty 
residences where they are happy, with ``happy'' being defined as having at least
half of the  neighbours coming from the own group, and ``unhappy''  meaning that
the majority of neighbours comes from the other group. This version leads to 
small clusters \cite{kirman,solomon} but not to large domains as in Harlem, 
New York City. Removing and replacing some people randomly \cite{jones} helps,
and one also gets ``infinitely'' large domains if happy (unhappy) people move 
to other places where they are happy (unhappy) \cite{kirman}. More plausible is 
a finite temperature or tolerance $T$ \cite{ortmanns,schulze,solomon}, where people
sometimes also move from happy to unhappy places, e.g. because they got a new
job elsewhere. 

While then the domains in the Schelling model at positive temperature seem to 
grow towards infinity \cite{solomon}, is is simpler to achieve the same
aim in the well-known two-dimensional Ising model, with or without conserved 
number of people within each group. 
Flipping a spin then means that a person from one group leaves
town and is replaced by a person from the other group, as simulated already
by \cite{jones}. Unfortunately, sociologists ignored for decades the Ising 
model (and also barely cited \cite{jones}) while physicists until 2000 mostly 
ignored the Schelling model.

\section{From Ising to Schelling}
This section moves fromt he simple Ising model with Glauber dynamics through
various intermediate stages up to the complicated Schelling model, in order to 
check when we get domains as large as the lattice size, and when only 
clusters much smaller than the lattice size. We always employ a $79 \times 79$
square lattice and check also positive temperatures $T < T_c$. Figures and 
Fortran program are given elsewhere \cite{mueller}.

The Glauber dynamics at $T>0$, as opposed to the one at $T=0$, is well known to 
give first two domains, and finally one domain overwhelms the other. For 
all following models, the number of people belonging to the two groups remains
(roughly) constant, and thus we never get only one domain. In Kawasaki dynamics
people of group A exchange residences with people of group B, and two domains
are formed for $T > 0$ both in the case of nearest-neighbour exchange and in the
socially more realistic case of exchanges at arbitrary distances. 

Closer to Schelling are variants without an energy and with merely a state of 
happiness: A resident is happy if surrounded by a majority of the own group,
unhappy inside a majority of the other group, and neutral if half of the 
neighbours belong to group A and half to group B (not counting empty sites).
Unhappy people try to exchange residences with unhappy people from the other 
group, such that they both become happy. Happy people do not move. Neutrals 
are treated according to  three variants: 
\begin{itemize}
\item a) Neutrals are treated as happy; thus only unhappy people exchange sites.
This option was taken by Schelling.
\item b) Neutrals are treated as unhappy and thus try to exchange, also with 
neutral places such that nobody gains.
\item c) Only unhappy people exchange, also with neutral places; both gain.
\item d) Unhappy and neutral people exchange only with unhappy people; both 
gain.
\end{itemize}

For $T = 0$, first we check neighbours in a fixed order: none of these variants
gives domains. Second we select one of the four neighbours randomly: only the 
socially unrealistic variant b gives domains.

In the version for $T > 0$, we follow \cite{solomon} and use Glauber-type 
probabilities with pseudo-energies $+1, 0, -1$ for unhappy, neutral and happy, 
respectively. It does not matter whether an A is surrounded by three B and one
A, or by four B: A is equally unhappy in both cases. Again we get domains only
for variant b, both for exchange with a random nearest neighbour, and for 
exchange with a random site anywhere. 

Schelling did not let two people exchange residences directly, but only let
one person at a time move into an empty residence. When we assume 20 percent 
of the sites to be empty and allow to move anywhere in the lattice we get
no domains, neither for $T=0$ not for $T > 0$.

Schelling let unhappy people move into the nearest empty place where they 
would not be unhappy. This prevents domains at $T = 0$ \cite{kirman,solomon}. 
However, if people also make neutral moves where their degree of happiness
remains constant, domains are formed at $T = 0$ \cite{kirman}. At positive
temperature, domains are formed \cite{solomon} without the unrealistic 
requirement that people make the moving effort without gain.

Finally, avoiding a positive temperature but producing random noise in a 
different way Dethlefsen and Moody \cite{jones} found large clusters (but 
according to our simulations no infinite domains), while Jones \cite{jones} 
found domains. 

Thus randomness, e.g. through a positive temperatures, produces large ghettos
(domains), while without such randomness and without the unreasonable 
requirement to move without any improvement, only small clusters and no real
ghettos are formed. The complications built in to move from Ising to Schelling
had only minor influence.
 
\section{Modified Ising model}
In the spin 1/2 Ising model with Glauber kinetics on the square lattice, we interpret
the two spin orientations as representing two groups of people, printed as
black and white in our later figure. Nearest neighbours
are coupled ferromagnetically, i.e. people prefer 
to be surrounded by others of the same group and not of the other group. Starting
with a random initial distribution of zero magnetisation (= number of one group minus
number of the other group), we check if ``infinitely" large domains are formed.
It is well known that they do so for $0 < T < T_c$ where $T_c = 2.269$ is the critical
temperature in units of the interaction energy. (More social interpretation
and explanation of the Ising model is given e.g. in \cite{solomon}.)

\begin{figure}[hbt]
\begin{center}
\includegraphics[angle=-90,scale=0.30]{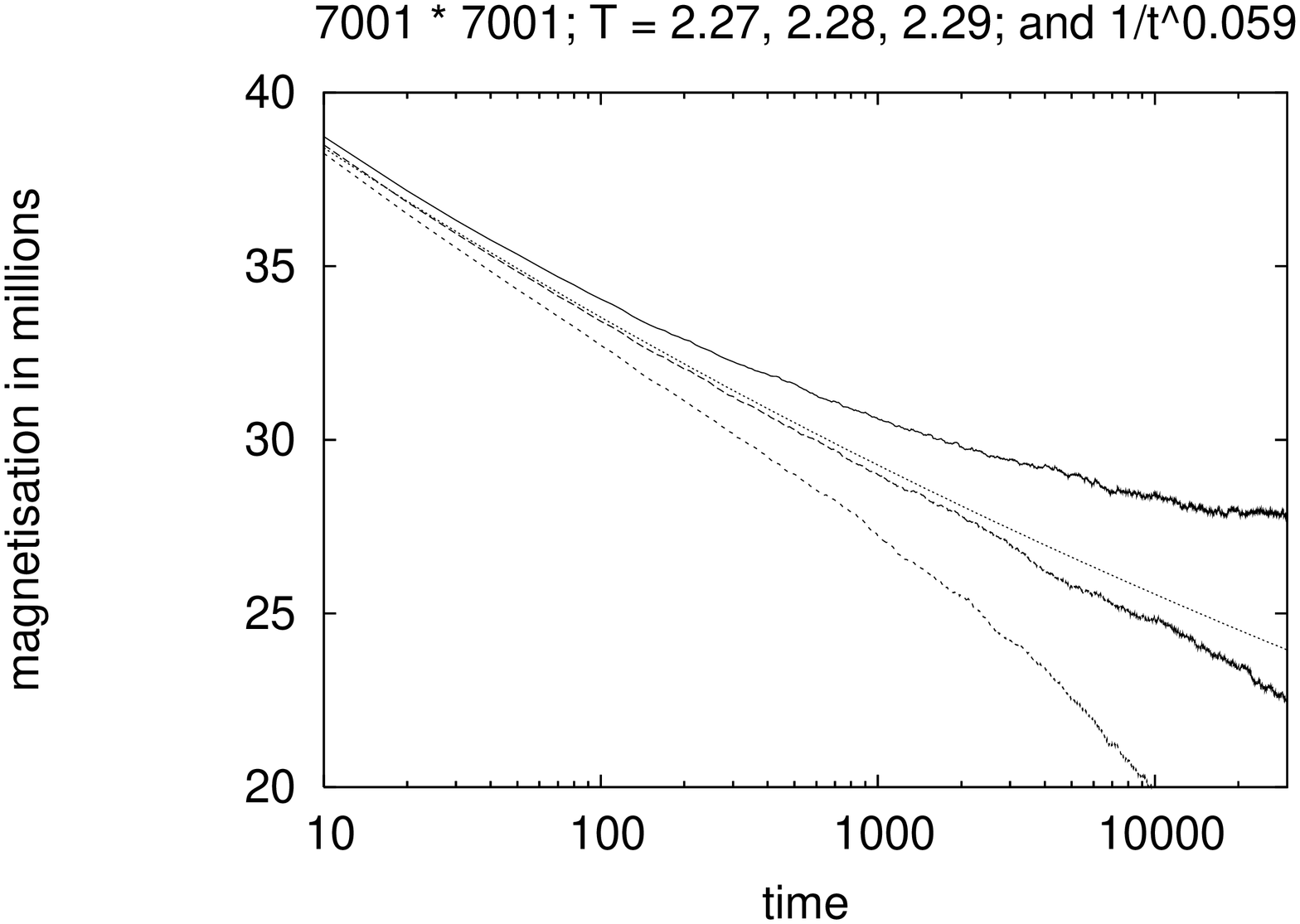}
\includegraphics[angle=-90,scale=0.30]{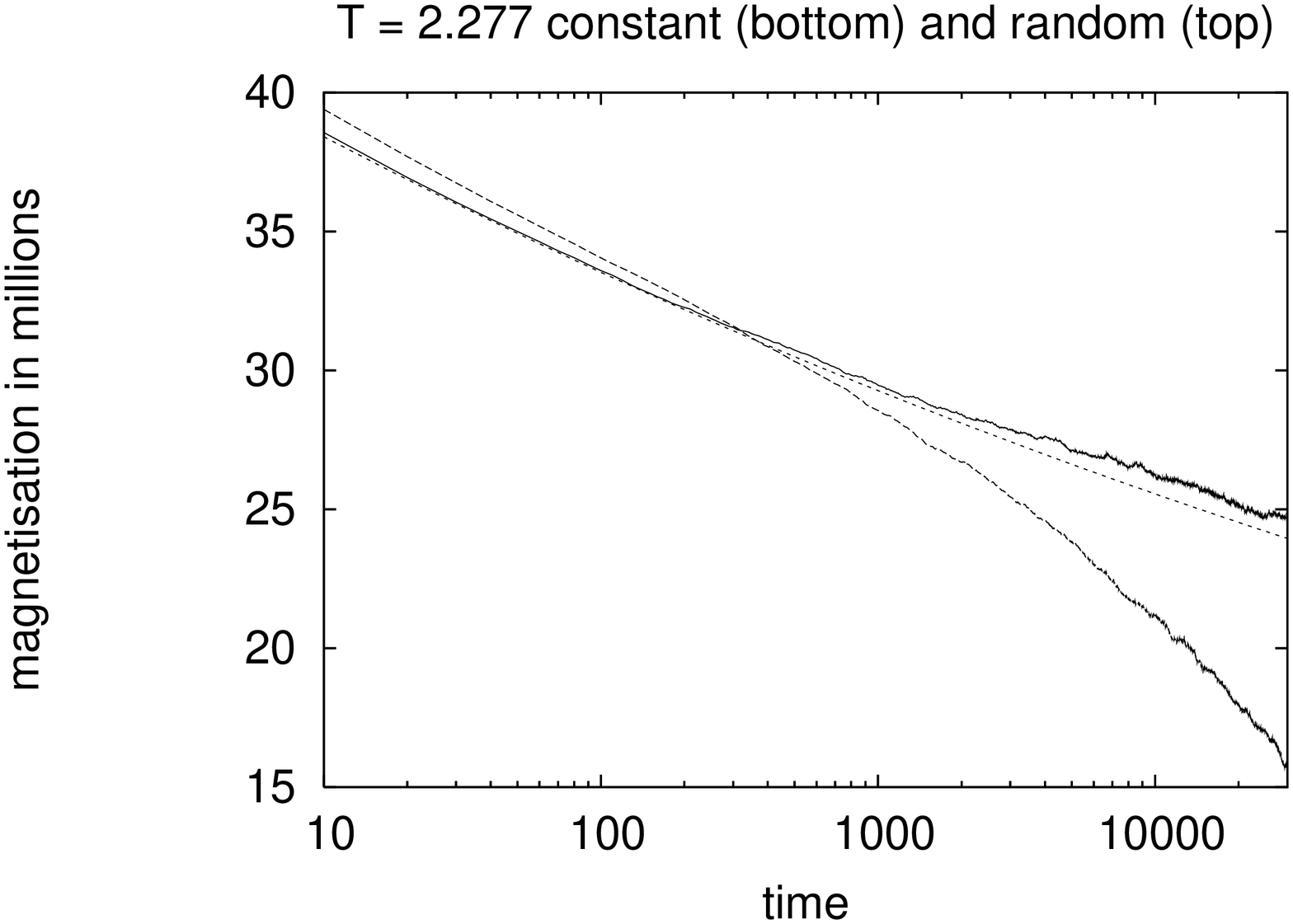}
\end{center}
\caption{
Decay of magnetisation if we start with all spins up: the critical
temperature seems to be between 2.27 and 2.28, i.e. between the upper and the
central curve. The smooth line corresponds to the theoretical variation in the
critical standard Ising model. (No feedback).
}
\end{figure}

\begin{figure}[hbt]
\begin{center}
\includegraphics[angle=-90,scale=0.31]{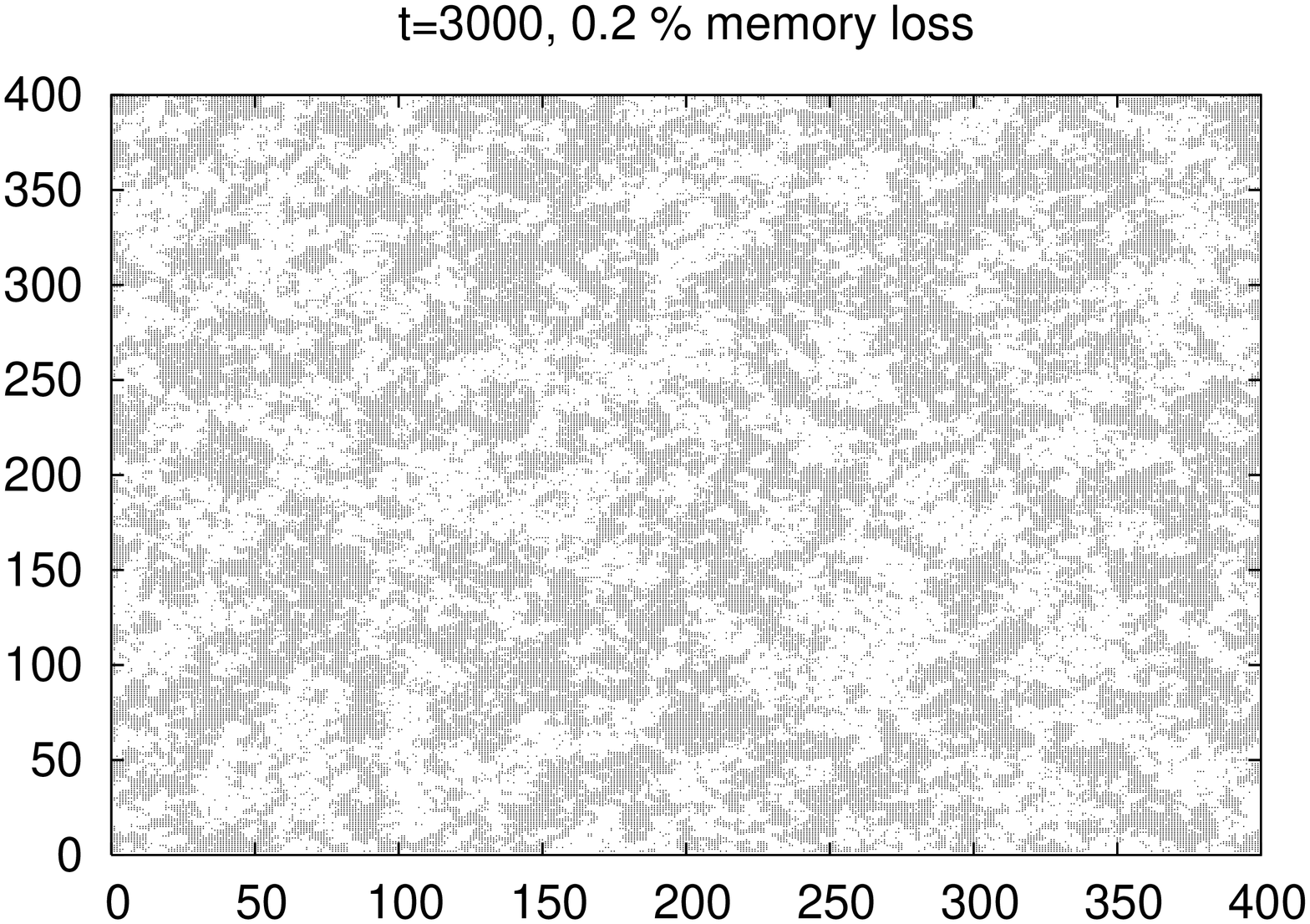}
\includegraphics[angle=-90,scale=0.31]{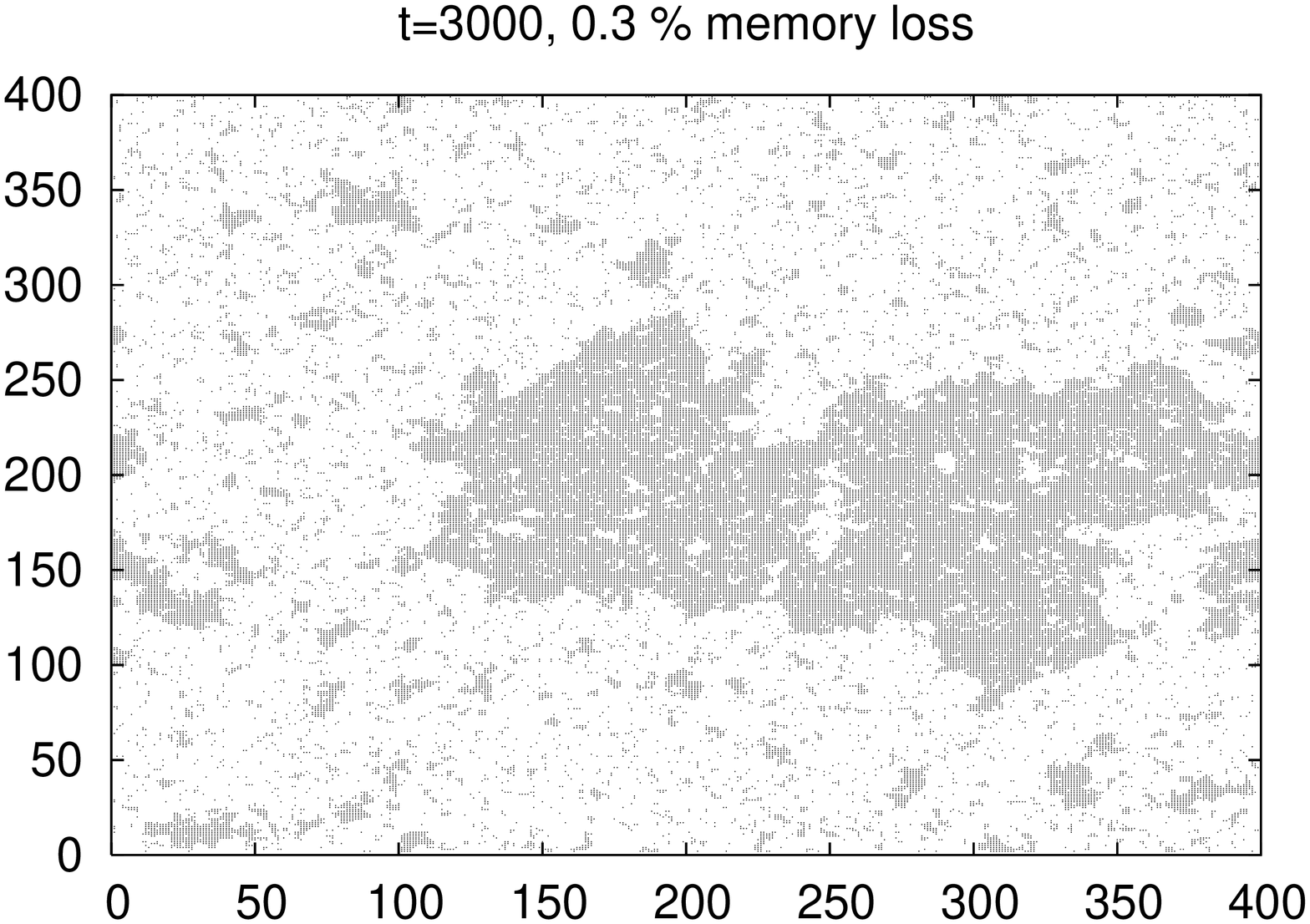}
\end{center}
\caption{
Small clusters (part a) and large white domain (part b)  with feedback. The
average initial temperature is 1.5. 
}
\end{figure}

\begin{figure}[hbt]
\begin{center}
\includegraphics[angle=-90,scale=0.3]{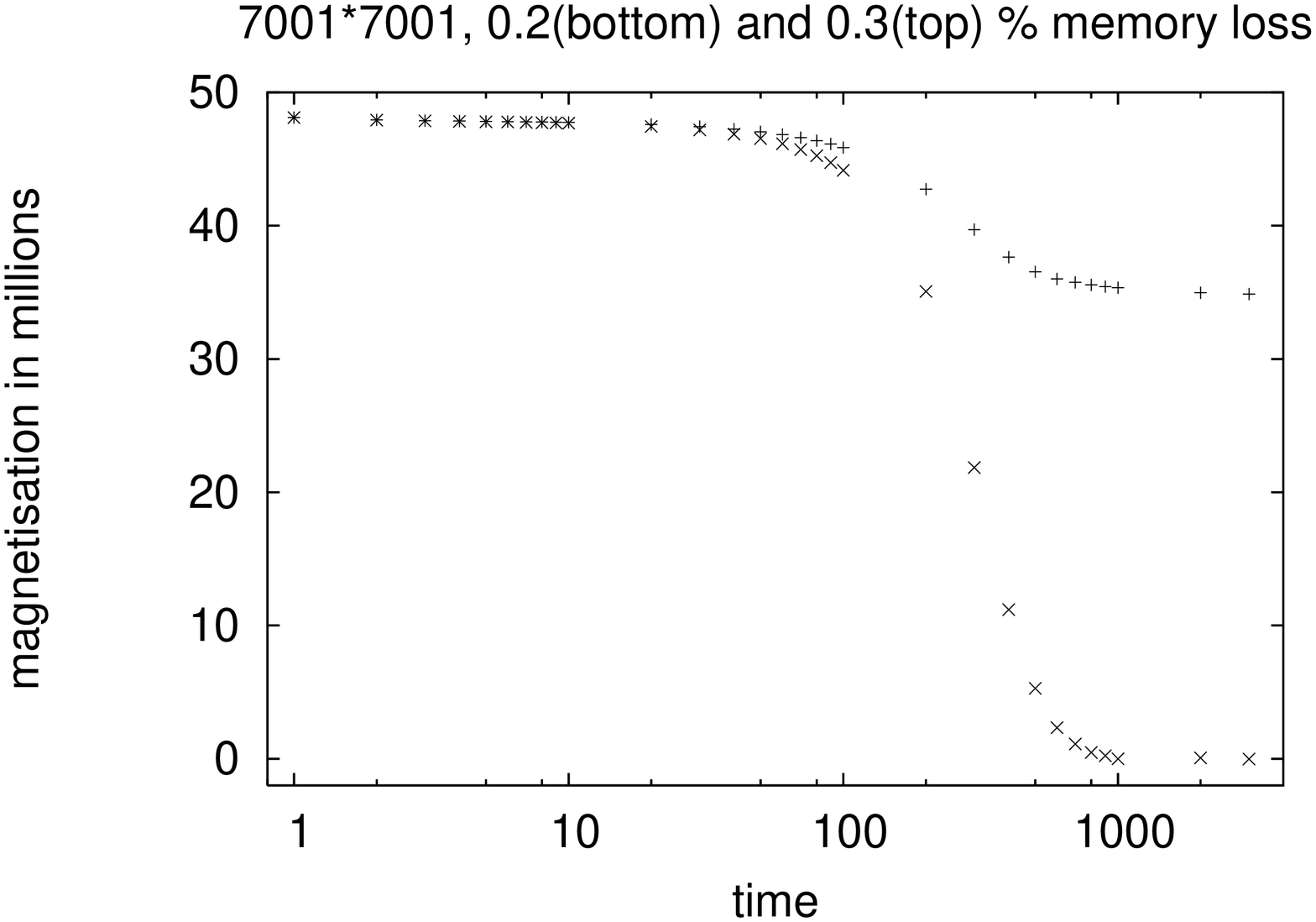}
\includegraphics[angle=-90,scale=0.3]{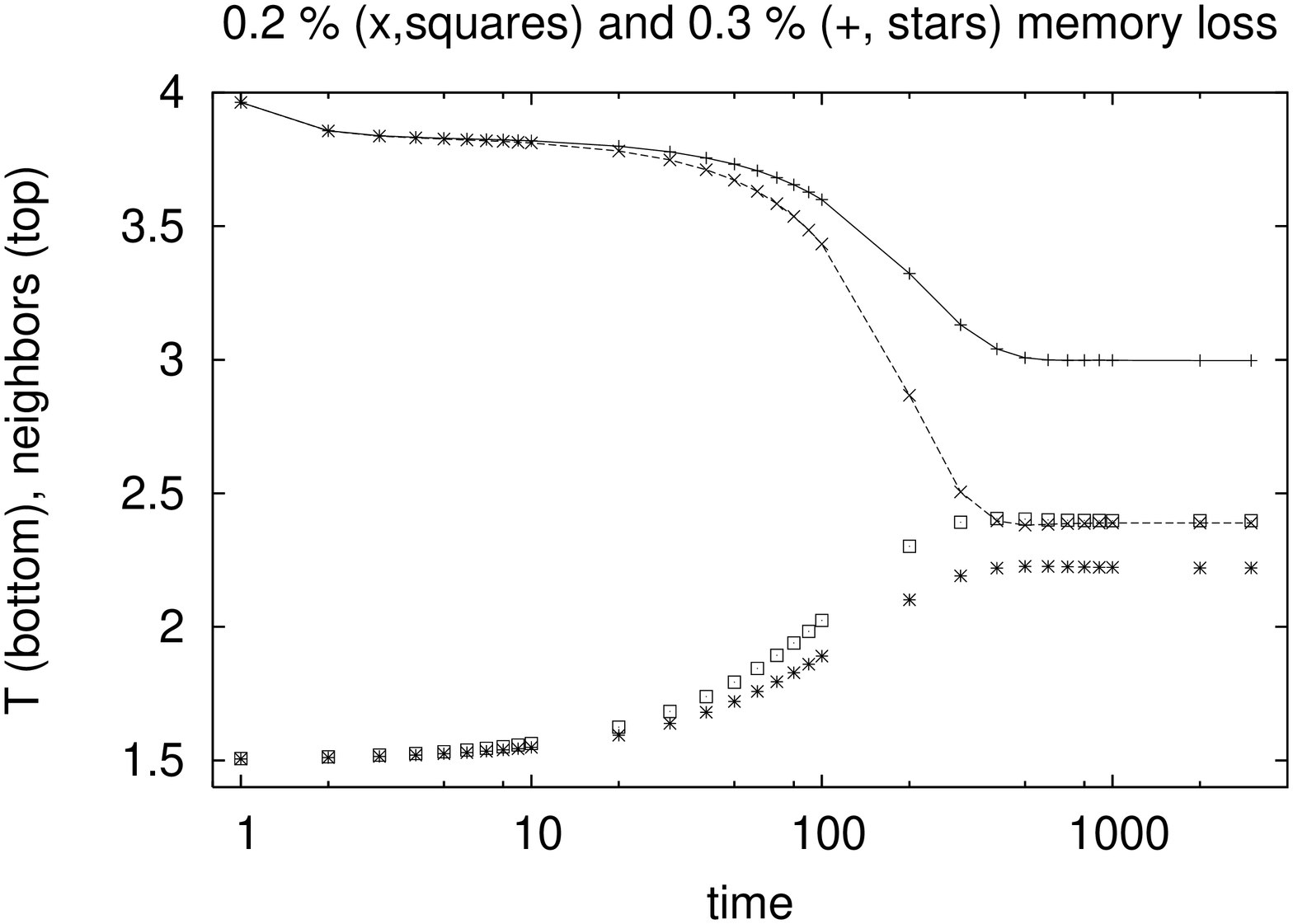}
\end{center}
\caption{
a: Magnetisation versus time with feedback and with cooling rate (forgetting) of
0.2 and 0.3 percent per iteration.
b: self-organisation of the 
temperature and the resulting average composition of the immediate 
neighbourhood of each site.
}
\end{figure}

Inhomogeneous temperatures are now introduced into the standard Ising model in order
to take into account that different people are different; some are more tolerant and
some less. Thus initially every site gets a temperature selected randomly between 
50 and 150 percent of the average temperature. Surprisingly, very little is changed 
in the results: The critical value for the average temperature increases 
from 2.269 to about 2.277, as
seen in Fig. 1. For lower temperatures large domains grow towards infinite size,
for higher temperature they remain finite (not shown.) 

(In these simulations the temperature is associated with a site. If a spin flips, 
that means if a person moves away and somebody of the opposite group moves from
another city into this residence, the new resident takes over the old value for
the tolerance or temperature $T$. If instead the new resident gets a new, 
randomly selected $T$, then the transition increases to $\langle T \rangle 
\simeq 2.64$, not shown.)

More interesting is the case of a feedback \cite{sornette,bornholdt}. We assume that
people are aware of the dangers of segregation, but still (in contrast to
\cite{sornette,bornholdt} but in the spirit of \cite{schelling,jones}) rely only on
their neighbourhood, not on any mass media or government orders. Thus if all four 
neighbours of one person belong to the same group that person increases the own
$T$ by 0.01; if all four neighbours belong to the other group, the local $T$ 
is decreased by the same amount 0.01. In order to prevent $T$ to move towards 
infinity, one also forgets at every iteration a small amount of tolerance, i.e.
$T$ is decreased by a few tenths of a percent independent of the neighbourhood.

Fig.2 shows configurations for 0.2 and 0.3 percent forgetting (= cooling) rate. 
For the lower rate, we see finite clusters; after a ten times longer simulation
the picture looks about the same. For the higher cooling rate a large white domain
has formed; after a ten times longer simulation the remaining black domain has become
much smaller. More quantitatively, we start from all spins up and monitor the 
magnetisation versus time in Fig.3a. For a rate of 0.2 percent the magnetisation 
drops to nearly zero, for a rate of 0.3 percent it stays rather close to its 
initial value. For the same simulations, Fig.3b shows the increase with time of 
the average $T$ (squares and stars) and the decrease of the number of neighbours
(+ and $\times$), for these two rates. Here the difference between the two
rates is less drastic. (The upper part shows the average number of neighbours 
of the same group, minus the average number of neighbours of the opposite group.
For each single site this difference can vary between --4 and +4.) The phase 
transition is near a rate of 0.29 percent memory loss (not shown). 
Not much is changed (not shown) if the local temperature $T$ shows a heat 
conduction, i.e. if people learn (in-)tolerance from their neighbours by taking 
at each iteration the average $T$ of their four neighbours.

\section{Summary}

The complications of the Schelling model are not necessary; the earlier standard
Ising model gives similar results. A new feedback between the composition
of the neighbourhood and a local temperature gives a self-organised temperature
which may be on one or the other side of a phase transition between 
segregation and mixing. 

We thank W. Alt and V. Jentsch (complexity center at Bonn university) for 
suggesting inhomogeneity, heat conduction and feedback.

\end{document}